# Benchmark Test of CP-PACS for Lattice QCD[*]

T. Yoshié

Center for Computational Physics,
University of Tsukuba, Ibaraki 305, Japan

## Abstract

The CP-PACS is a massively parallel computer dedicated for calculations in computational physics and will be in operation in the spring of 1996 at Center for Computational Physics, University of Tsukuba. In this article, we describe the architecture of the CP-PACS and report the results of the estimate of the performance of the CP-PACS for typical lattice QCD calculations.

## 1  CP-PACS Project

The CP-PACS project is a collaboration of computer scientists and physicists[1] to develop a massively parallel computer dedicated for large scale scientific calculations and to do research with it in several areas of computational physics, in particular on lattice QCD. The CP-PACS with 1024 nodes is scheduled to be completed by March 1996 and we plan to start physics calculations in 1996. The project is being carried out in close collaboration with Hitachi Ltd.

In this article, we first describe the CP-PACS computer briefly and then report the results of benchmarks for performance, focussing on calculations of light hadron spectroscopy with Wilson quarks.

For preceding reports on the CP-PACS, see ref.[2].

---

[*] talk presented at "QCD on Massively Parallel Computers" held at Yamagata, Japan, on March 16-18, 1995.

# 2 CP-PACS Computer

## 2.1 Nodes and Network

The CP-PACS is a MIMD parallel computer with distributed memory and distributed disks. It consists of 1024 processing units (PU) and 64 I/O units (IOU). The nodes are interconnected by a flexible network which we call Hyper-Cross-Bar network. In fig.1 is schematically illustrated the network configuration. 1024 PUs are arranged in a three dimensional $8 \times 16 \times 8$ array. Crossbar switches are placed parallel to the three axes. To connect the PU to the crossbars and the crossbars in different directions themselves, a 4 by 4 crossbar switch called Exchanger is placed on each node. Data transfer from a node to any other node can be made in at most three steps. Though-put per each connection is expected to be more than 200 MByte/sec. 64 IOUs are arranged in a two dimensional $8 \times 8$ array and are attached at the end of the PU array to join the same network of the PUs.

Schematic diagrams of the PU and the IOU are shown in fig.2. Each PU has one super scalar RISC chip of a peak speed of 300 MFlops with an enhancement of pseudo vector processing (PVP). The PVP is a unique feature of the CP-PACS and will be described in the next subsection. Local Storage of 64 MByte is constructed from several banks of DRAM. Storage Controller supplies data in main memory to the processor or transmit them to the network through the Network Interface Adapter. Each IOU has the same components of the PU together with Bus Adapters and IO Adapters. A RAID5 disk with capacity of 8.4 GByte is attached to the IO Adapter through SCSI-II interface.

## 2.2 PVP-SW

It is well known that vector processing is suitable for a wide class of large scale scientific calculations. Therefore we have chosen a strategy to incorporate vector processing with a commercial RISC architecture. (We have adopted Hewlett Packard PA-RISC.) The enhancement is called PVP-SW[3], pseudo-vector processing based on slide windowed registers.

First we need many floating point registers to handle complex operations. However, instructions of PA-RISC can specify only 32 registers. "Slide windowed floating point registers" illustrated in fig.3 is a solution to handle many physical registers without serious change of instruction set. The 32 registers specifiable with the PA-RISC is logically divided into two parts: global registers and local registers. The global registers are always assigned at the top of the physical registers. On the other hand, we make that the local registers can slide along the rest of the physical registers. Totality of the local registers at a particular position together with the global registers is called a window. Arithmetic instructions can be executed only on an active window and therefore no change of instruction set for arithmetic operations is necessary. User can handle all physical registers by changing the active window successively.

The slide windowed registers are utilized to hide a large memory access latency. We introduce new instructions called Preload and Poststore. The Preload instructions can fetch data from memory to any physical registers and the Poststore instructions can store data from any physical registers. After the issue of the Preload/Poststore instruction, processors do not wait the data and proceed to the next instruction. If the number of registers is enough, we can schedule the Preload instructions in such a way that data already reside in registers when the registers become active.

A high bandwidth between main memory and processor is also necessary for vector processing. As is well known, high performance of RISC chip for arithmetic operations highly depends on the existence of data cache. However, in most of lattice QCD calculations, data size is larger than cache size and data access is done with low locality. This is the reason for the necessity of high bandwidth between main memory and processor. For this purpose, the PU has the Storage Controller to pipeline the memory with multiply interleaved banks. Bandwidth is one double precision data per one machine cycle.

Because PA-RISC is a super scalar chip which can execute a multiplication and an addition in one machine cycle, we can issue Preload or Poststore, Multiplication and Addition at each machine cycle without stall owing to these enhancement. The three types of instructions are well balanced in most of QCD calculations. Therefore we can expect high performance for QCD calculations on the PVP-SW chip.

## 3 Lattice QCD Benchmarks

We discuss how the CP-PACS architecture works for QCD calculations and make an estimate of performance, focussing on calculations of light hadron spectroscopy with Wilson quarks. For the benchmarks, we assume the CP-PACS has 2048 PUs and 128 IOUs. (A new funding request to scale up the machine to 2048 nodes is being made. However, the final decision will be made early in 1996.)

### 3.1 Wilson Matrix Multiplication on 1 PU

We have extensively analyzed the performance for Wilson matrix multiplication (MULT), because it is the most time consuming in spectroscopic studies. The three types of instructions Preload/Poststore, Multiplication and Addition are well balanced in MULT, and therefore the CP-PACS exhibits high performance. Hereafter we show the performance in terms of efficiency which is defined by the ratio of the sustained speed to the theoretical peak speed or equivalently the ratio of twice the number of floating point operations to the number of machine cycles (MC).

The MULT can be symbolically written as 1) $h = g - \kappa \sum_\mu p$ and 2) $p = (1 \pm \gamma_\mu)Uq$, where $\kappa$ is the hopping parameter, $U$ is a link variable and $h, g, p$ and $q$ are $3 \times 4$ matrices. The calculation of 2) in spatial directions for each spinor index requires 30 loads, 6 stores together with 36 multiplications and 36 additions. Because of this balance of the three types of instructions, we can perform it in 36 MC. The efficiency is 100 % for this part. For a similar calculation in the time direction and the calculation of 1), the instructions are less balanced and machine cycles can not be filled up with Multiplications and Additions. The efficiently falls to 90% because of the unbalance.

In collaboration with Hitachi Ltd., we have developed a hand optimized code including PVP-SW features. The code takes account of window switches, address calculations and all the other restrictions due to instruction processor and storage controller. Assuming that the number of registers is enough to hide memory access latency and that the lattice size is infinite, we find that the efficiency turns out to be 75%.

Next we consider a finite volume effect on the performance. Coding with PVP-SW needs preload-phase before the body of the do loop to fill up the registers with data and poststore-phase after the do loop to sweep out the data. This is the origin of the finite volume effects. We subdivide the lattice into sub-lattices only in the spatial directions on a three dimensional PU array. Therefore, the most inner do loop, which should be as long as possible, is in the temporal direction. Fig.4-a) shows the efficiency versus the temporal extension of the lattice. The figure shows that the fall of the efficiency is at most 2% even on a small lattice such as a $10^4$ lattice.

Fig.4-b) shows the efficiency versus the number of registers for the case of a $64^4$ lattice. Memory access latency depends on memory access pattern due to the bank memory structure. In order to estimate the bank conflict effect, we have analyzed the code using a memory simulator which incorporates the actual structure of memory. We find that the fall of the efficiency is less than 1%, if the number of registers is 100 or more. Since our processor has 128 registers, the expected performance is 74 %.

## 3.2 Wilson quark R/B MR solver

Wilson quark red/black (R/B) minimal residual (MR) solver is coded as follows. We prepare the residual vector $r_e$ on even sites in advance. First we multiply odd-even and even-odd Wilson matrices to calculate the search direction: $p_e = D_{eo}D_{oe}r_e$. Then we calculate the $\alpha = (p_e, r_e)/(p_e, p_e)$ and replace the solution vector $x_e$ and the residual vector $r_e$; $x_e = x_e + \alpha r_e$, $r_e = r_e - \alpha p_e$. In addition we calculate the residual sum of squares $(r_e, r_e)$ for convergence check. We have estimated execution times for these operations.

Table 1: Estimate of the number of iterations to solve the quark propagator for each color and spinor by the red/black minimal residual method

| $m_\pi/m_\rho$ | 0.9 | 0.8 | 0.7 | 0.6 | 0.5 | 0.4 | 0.3 | 0.18 |
|---|---|---|---|---|---|---|---|---|
| #iter | 200 | 200 | 360 | 630 | 1060 | 1830 | 3470 | 10130 |

Before each multiplication of the Wilson matrices, we need to transfer the data on even or odd sites and on the boundary of sub-lattices. We assume we do not overlap the data transfer and the arithmetic operations, though it is possible on the CP-PACS. We assume that the network latency is 5 micro second. We consider two cases of the network through-put, 150 MByte/sec and 300 MByte/sec. We also need a global sum when calculating the $\alpha$ and the $(r_e, r_e)$. Cascade sum is assumed for the global sums.

Fig. 5-a) shows the ratio of the data transfer time to the total execution time in the MR loop one iteration. For the case of 300 MByte/sec through-put, the data transfer account for 10 % on a $64^4$ lattice and amount to 25 % on a $32^4$ lattice. On a smaller lattice such as a $16^4$ lattice, half or more of the execution time is spent for the data transfer. Fig.5-b) shows the efficiency versus the lattice size. For the case of 300 MByte/sec through-put, the efficiency decreases from 74 % to 66 (57) % on a $64^4$ ($32^4$) lattice. Actual execution time for one iteration is 0.063 second for a $64^4$ lattice.

### 3.3 Quenched QCD

The spectrum calculation in quenched QCD consists of configuration generation, quark propagator calculation and construction of the hadron propagators. We have estimated the execution time for each part, taking 6 hopping parameters which correspond to the $m_\pi/m_\rho$ of 0.9 to 0.4 in steps of 0.1.

The estimate of the execution time for quark propagator calculation crucially depends on the estimate of the number of iterations we need to solve the propagators. Because the condition number of the Wilson matrix is approximately proportional to the inverse of the quark mass, the number of iterations is approximately proportional to the inverse quark mass. To estimate the number of iterations, we have used the proportional constant estimated using the data from QCDPAX collaboration. (Their stopping conditions are $\sqrt{\sum |r|^2/3 \cdot 4 \cdot V} < 10^{-9}$ and $|r/x| < 0.03$ for each degree of freedom.) In table. 1 are quoted the number of iterations we have adopted for the estimate of the execution time.

Disk I/O is another feature of the quark propagator calculation. We are going to save all of the six propagators to disks, and after the propagator of the lightest quark is calculated, all the propagator will be read for the construction of the hadron propagators. This is necessary in order to calculate observables for all the combinations of quark masses. We have estimated the I/O time to write and read each propagator once, using the specification of the RAID5 disks we adopt. For the case of 300 MByte/sec network throughput, the ratios of the execution times for the floating point operations, the data transfer and the disk I/O are 62%, 8% and 30% ,respectively, on a $64^4$ lattice. We notice that the I/O is relatively heavy for quenched QCD calculations. Actual execution time to calculate propagators on one configuration is 1.27 hour for a $64^4$ lattice.

Table 2: Execution time (days) of quenched QCD run described in the text.

| Size $L^4$ | 500 conf. | 500 $\times(64/L)^4$ conf. |
|---|---|---|
| $16^4$ | 0.34 | 87.37 |
| $32^4$ | 3.41 | 54.55 |
| $48^4$ | 15.70 | 49.61 |
| $64^4$ | **47.63** | 47.63 |
| $80^4$ | 113.65 | 46.55 |

The execution time for configuration generation is estimated by counting the number of floating point operations. We assume five hit pseudo heat bath algorithm with three SU(2) sub-matrices. Our estimate of the total number of floating point operations equivalent is 5700 per link. Using an estimate of the number of floating point operations for the R/B MR one iteration, we find that the heat bath 1 sweep is approximately 15 minimal residual iterations equivalent. This estimate corresponds to 14 $n$-sec per link update. The execution time of the hadron propagator calculations is estimated using data from QCDPAX collaboration. Assuming that each configuration is separated by 3000 sweeps, we estimate that 35% of the total time is used for the configuration generation, 10% for the hadron propagator calculations, and 55% for the quark propagator calculations.

Table 2 shows the total execution time for 500 configurations. We need 48 days for a $64^4$ lattice. It is well known that the number of configurations we need to obtain results with similar statistical errors strongly depends on the volume of the lattice. To take account of this property, we have tabulated in table 2 the execution time for configurations of 500 times the volume ratios, assuming that the number of configurations we need are proportional to the inverse of the volume. From this point of view, the volume dependence of the execution time is marginal, except for a small lattice such as $16^4$. Therefore the total execution time for quenched QCD run is approximately 50 days times the number of combinations of the $\beta$ and the lattice size, assuming that the statistics is 500 configurations for a $64^4$ lattice and equivalent for the others.

Table 3: Execution time (days) of full QCD run described in the text.

| Lattice Size | | $m_\pi/m_\rho$ | | | | |
| --- | --- | --- | --- | --- | --- | --- |
| $L_s$ | $L_t$ | 0.7 | 0.6 | 0.5 | 0.4 | 0.3 |
| 16 | 32 | 1.3 | 2.3 | 3.9 | 6.7 | 12.7 |
| <u>32</u> | <u>64</u> | **29** | **51** | **85** | **147** | 279 |
| 48 | 64 | 135 | 236 | 397 | 686 | 1300 |

If we take a lattice with the lattice spacing of 0.05 fm and the size of 64, the physical lattice size is 3.2 fm. Roughly speaking, this lattice is 1.5 times large and 1.5 times fine compared with a typical lattice of today's calculations. With the physical lattice size held fixed, we expect we can carry out simulations for three values of the lattice spacing to take the continuum limit, within 150 days.

## 3.4 Full QCD

For full QCD simulations, we have estimated the execution time to generate configurations. We take Hybrid Monte Carlo algorithm with molecular dynamics step size of $0.02 \times 8/L_s$. ($L_s$ is the spatial lattice extension.) We have estimated the execution time for only the inversion of pseudo fermion variables. The number of iterations is estimated by the same method used for quenched QCD. The stopping condition is changed to $\sqrt{\sum |r|^2/3 \cdot 4 \cdot V} < 10^{-7}$.

Table 3 shows our estimate of the execution time to generate 500 configurations, each of them being separated by 5 units of simulation time. Data are quoted for each $m_\pi/m_\rho$. The situation is severe for a $48^3$ lattice. For a $32^3 \times 64$ lattice, taking the $m_\pi/m_\rho$ of 0.7 to 0.4 in steps of 0.1, we estimate the total execution time is 310 days for each $\beta$.

## 4 Summary

We have estimated the execution time of simulations for light hadron spectroscopy with Wilson quarks on the CP-PACS assuming 2048 processing units and 128 I/O units. We have emphasized that the well balance of load/store, multiplication and addition is a noticeable feature of these calculations.

For quenched QCD, the total execution time is estimated to be 150 days, for six quark masses corresponding to the $m_\pi/m_\rho$ of 0.9 to 0.4 in steps of 0.1 at three $\beta$'s with the physical lattice size 3.2 fm fixed. (The largest lattice is a $64^4$ lattice with the lattice spacing of 0.05 fm.)

For full QCD, the execution time is estimated to be 310 days. It includes configuration generations at one $\beta$ for quark masses corresponding to the $m_\pi/m_\rho$ of 0.7 to 0.4 in steps of 0.1 on a $32^3 \times 64$ lattice. The estimated execution time seems too long for us to do calculations at several $\beta$'s. In order to resolve the situation, we are making an effort to search an efficient parallelization of fast inversion algorithms such as a conjugate residual method with incomplete LU decomposition which is efficient for vector machines.

I would like to thank the members of the CP-PACS project, especially Y.Iwasaki for valuable suggestions on the manuscript. This work is supported in part by the Grand-in-Aid of the Ministry of Education, Science and Culture (No.06NP0601).

# References

[1] The CP-PACS project is led by Y.Iwasaki. Other members are: S. Aoki, T. Boku, M. Fukugita, T. Hoshino, S. Ichii, M. Imada, N. Ishizuka, K. Kanaya, H. Kawai, T. Kawai, M. Miyama, S. Miyashita, M. Mori, Y. Nakamoto, H. Nakamura, T. Nakamura, I. Nakata, K. Nakazawa, M. Okawa, Y. Oyanagi, T. Shirakawa, A. Ukawa, M. Umemura, K. Wada, Y. Watase, Y. Yamashita, M. Yasunaga and T. Yoshié.

[2] Y. Oyanagi, Nucl.Phys. B (Proc.Supp.) 30 (1993) 299; Y. Iwasaki, Nucl.Phys. B (Proc.Supp.) 34 (1994) 78; A. Ukawa, Nucl.Phys. B (Proc.Supp.) 42 (1995) 194.

[3] H. Nakamura, H. Imori, K. Nakazawa, T. Boku, I. Nakata, Y. Yamashita, H. Wada and Y.Inagami, Proc. of International Conference on Supercomputing '93 (1993) 298; H. Nakamura, K. Nakazawa, H. Li, H, Imori, T. Boku, I. Nakata and Y. Yamashita, Proc. of 27th Hawaii International Conference on System Sciences, (1994) 368.

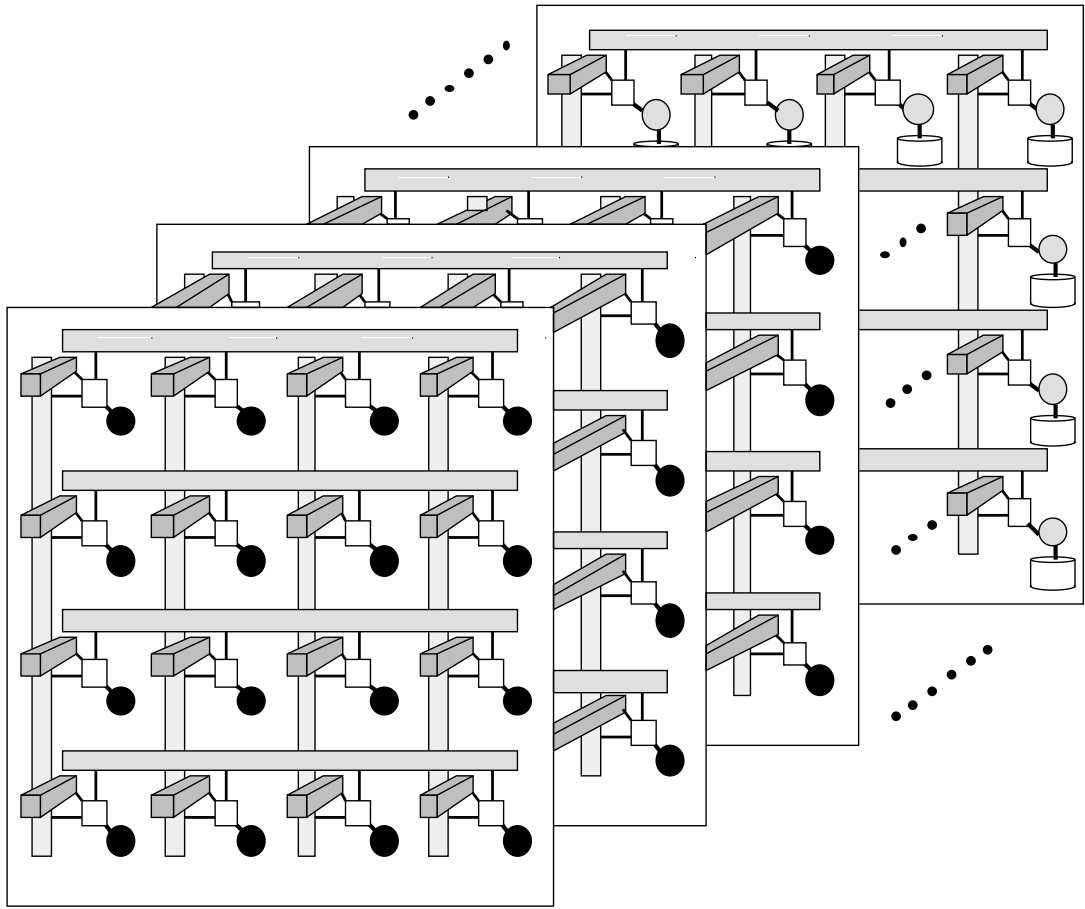

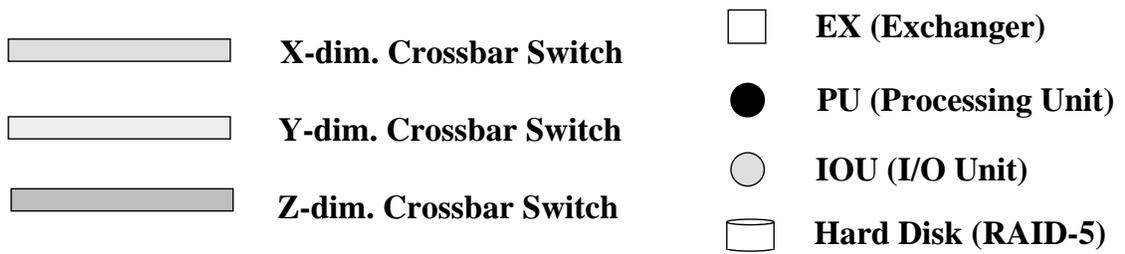

Figure 1: Illustration of the network configuration

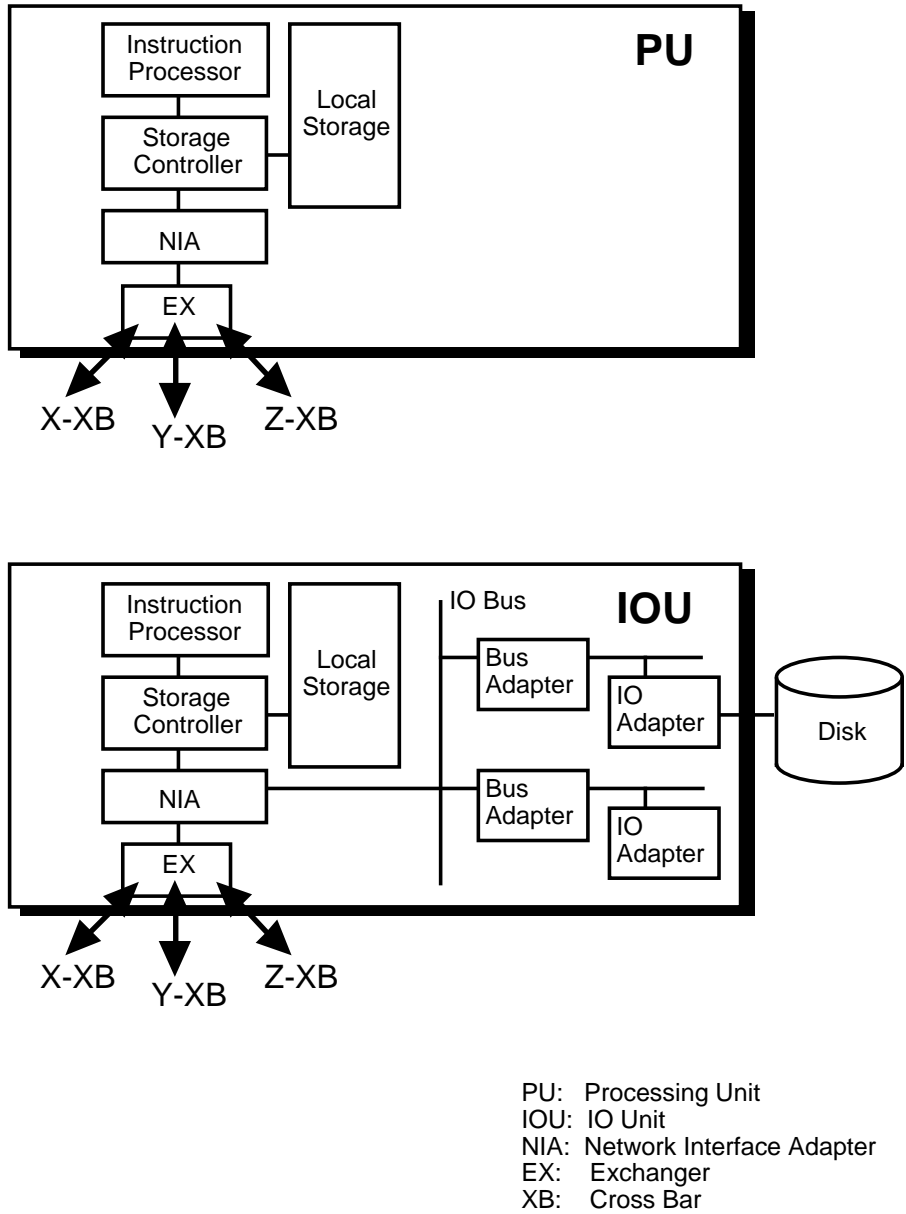

Figure 2: Schematic diagrams of PU and IOU

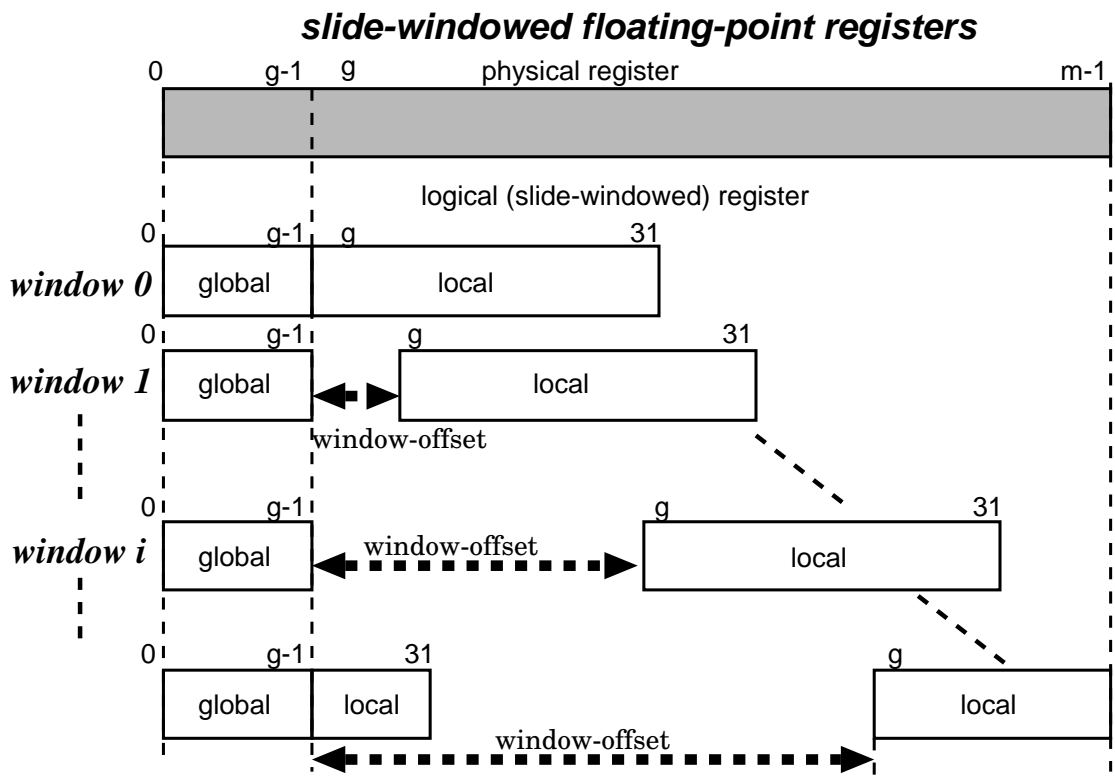

Figure 3: Structure of the slide-windowed floating-point registers

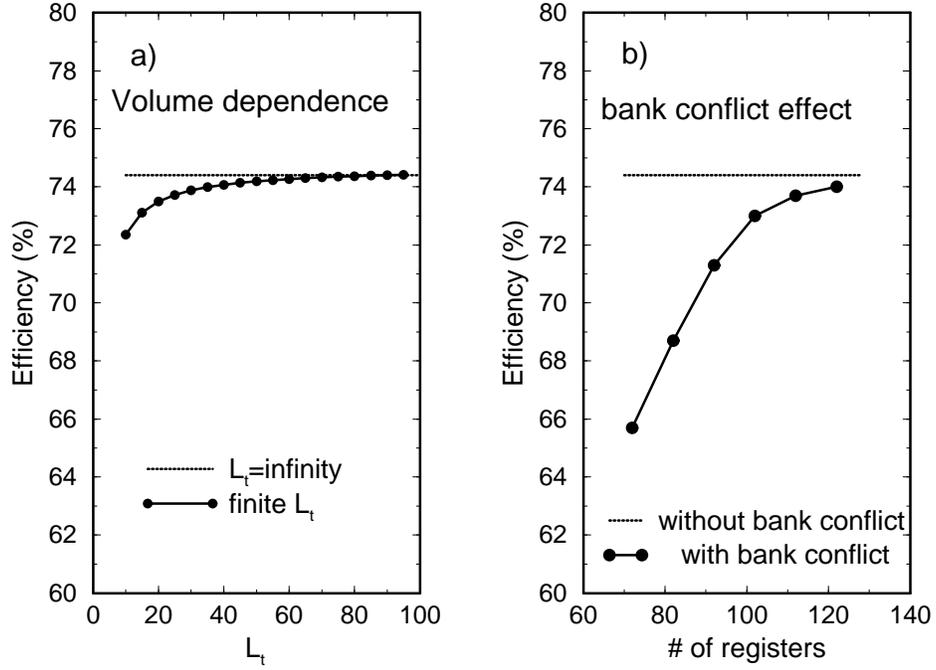

Figure 4: Efficiency of the CP-PACS for Wilson matrix multiplication

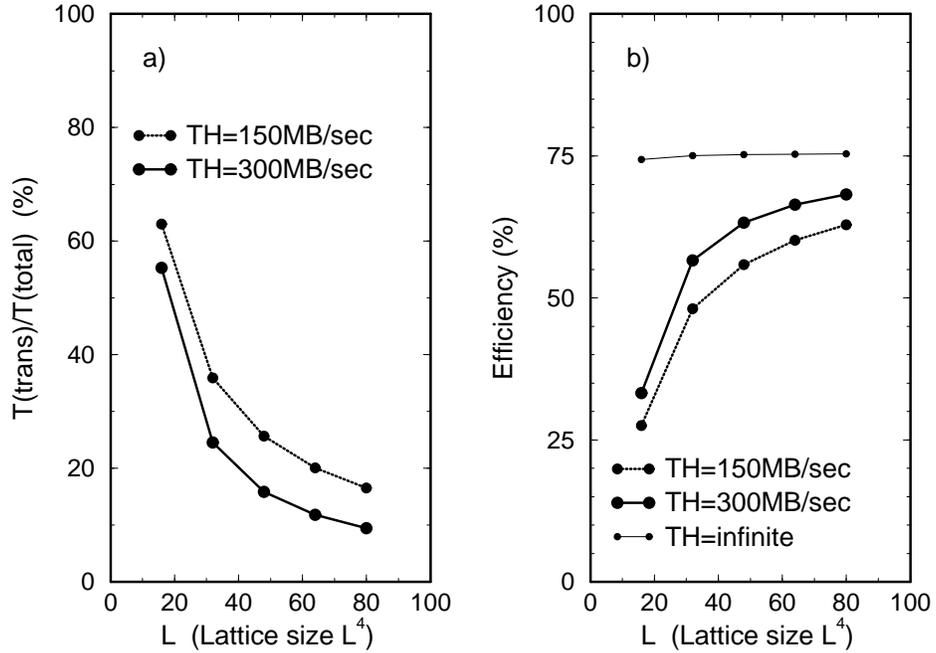

Figure 5: a) The ratio of the data transfer time to the total execution time and b) the efficiency for Wilson quark red/black minimal residual solver.